\input phyzzx.tex
\hfuzz 20pt
\font\mybb=msbm10 at 12pt

\def\Bbb#1{\hbox{\mybb#1}}

\def\bC {\Bbb{C}}

\def\bR{\Bbb {R}}

\def\fy {{\bf {y}}}
\def\fx {{\bf {x}}}
\def\fr {{\bf {r}}}
\def\fu {{\bf {u}}}
\def\fw {{\bf {w}}}
\def\fv {{\bf {v}}}
\def\bfomega{\omega\kern-7.0pt \omega}

\def\C{\mkern1mu\raise2.2pt\hbox{$\scriptscriptstyle|$}\mkern-7mu{\rm C}}

\def\log{{\rm log}}

\def\l{\lambda}
\def\a{\alpha}

\def\s{\sigma}
\def\pd{\partial_}
\def\b{\beta}

\def\g{\gamma}

\def\t {\theta}

\def\fp{{\bf{p}}}
\def\fk{{\bf{k}}}
\def\fu{{\bf{u}}}
\def\fv{{\bf{v}}}

\def\fab{{|{\bf{p}}_{AB}|}}
\def\fac{{|{\bf{p}}_{AC}|}}
\def\fbc{{|{\bf{p}}_{BC}|}}
\def\dilog{{\rm dilog}}

\def\b {\beta}

\def\a {\alpha}


\def\sdt{{\buildrel . \over \sigma}}

\REF\gibr {G.W.Gibbons and P.J.Ruback,{\sl
The motion of extreme Reissner-Nordstr\"om black
holes in the low velocity limit}, Phys. Rev.
Lett. {\bf 57} (1986) 1492.}
\REF\fere {R.C.Ferrell and D.M.Eardley,{\sl
Slow motion scattering and coalescence of
maximally charged black holes},Phys. Rev.
Lett. {\bf 59} (1987) 1617.}
\REF\shiraishi{K. Shiraishi, {\sl Moduli Space 
Metric for Maximally-Charged
Dilaton Black Holes}, Nucl. Phys. {\bf B402} (1993) 399.}
\REF\coles{ R. A. Coles and
 G. Papadopoulos, {\sl The Geometry 
of the One-dimensional
Supersymmetric Non-linear Sigma Models}, 
Class. Quantum Grav. {\bf 7} (1990)
427-438.}
\REF\gibbonsp{G.W. Gibbons, G. Papadopoulos  and K.S. Stelle, 
{\sl HKT and OKT Geometries on Soliton Black Hole Moduli Spaces},
 Nucl.Phys. {\bf B508} (1997)623; hep-th/9706207.}
 \REF\micha{J. Michelson and A. Strominger, {\sl Superconformal 
Multi-Black Hole
Quantum Mechanics}, JHEP 9909:005, (1999); hep-th/9908044.}
\REF\mss{A. Maloney, M. Spradlin and
 A. Strominger, {\sl Superconformal
Multi-Black Hole Moduli Spaces 
in Four Dimensions}, hep-th/9911001.
\hfill\break
R. Britto-Pacumio, J. Michelson, 
A. Strominger and A. Volovich,
{\sl Lectures on superconformal quantum 
mechanics and multi-black hole
moduli spaces}, hep-th/9911066.}
\REF\gutpap{J. Gutowski and G. Papadopoulos, {\sl The dynamics 
of very special black holes}, Phys.Lett. {\bf B472}:45-53, (2000);  
hep-th/9910022 .} 
\REF\gutpapb{ J. Gutowski and G. Papadopoulos, {\sl Moduli Spaces for 
Four-Dimensional and Five-Dimensional Black Holes}, 
 Phys.Rev. {\bf D62}:064023,2000: hep-th/0002242.}
 \REF\strqa{R. Britto-Pacumio, A. Strominger and A. Volovich,
 {\sl Two-Black-Hole Bound States}, JHEP 0103:050, (2001);  
hep-th/0004017.}
 \REF\strqb{R. Britto-Pacumio, A. Maloney, M. Stern and A. Strominger,
 {\sl Spinning Bound States of Two and Three Black Holes },
 hep-th/0106099.}
 \REF\claus{P. Claus, M. Derix, R. Kallosh, 
J. Kumar, P. Townsend and A. van
Proeyen, {\sl Black Holes and Superconformal Mechanics},
 Phys. Rev. Lett.
{\bf 81} (1998) 4553; hep-th/9804177.}
\REF\jade{J.A. de Azcarraga, J.M. Izquerido, 
J.C. Perez Buono and P.K. Townsend,
{\sl Superconformal Mechanics, Black Holes, and 
Non-linear Realizations}, Phys. Rev.
{\bf D59} (1999) 084015; hep-th/9810230.}
 \REF\aff{V. de Alfaro, S. Fubini and G. Furlan,  
{\sl Conformal Invariance in Quantum Mechanics}, 
Nuovo Cimento {\bf 34A} (1976) 569.}
 \REF\gibbonst{ G. W. Gibbons and P.K. Townsend, {\sl 
 Black Holes and Calogero Models},  
 Phys.Lett. {\bf B454}(1999) 187; 
 hep-th/9812034.}
\REF\twist{P.S. Howe and G. Papadopoulos, {\sl Twistor Spaces
for HKT Manifolds}, Phys. Lett. 
{\bf B379} (1996)80, hep-th/9602108.}

\Pubnum{ \vbox{ \hbox{}\hbox{} } }
\pubtype{}
\date{July, 2001}
\titlepage
\title{Three Body Interactions, Angular Momentum and Black
 Hole Moduli Spaces}
\author{J. Gutowski}
\address{Department of Physics\break Queen Mary College\break
 Mile End\break
London E1 4NS}
\andauthor{ G. Papadopoulos }
\address{Department of Mathematics\break King's College London \break  
Strand\break London WC2R 2LS}

\abstract {We investigate the dynamics of a pair of (4+1)-dimensional
black holes in the moduli approximation and with fixed  angular momentum. 
We find that  spinning black holes at small separations  
are described by the de Alfaro, 
Fubini and Furlan model. For  more than two black holes,
  we find an explicit expression for the three-body
interactions in the moduli metric by associating them
with the one-loop three-point amplitude 
 of a four-dimensional $\phi^3$
theory. We also investigate the dynamics of a three black hole system
in various approximations.}

\endpage
\pagenumber=2



\chapter{Introduction}

In the past two years there has been a revival in the interest of 
understanding the geometry of black hole moduli spaces in the
context of supergravity
following some earlier work in [\gibr, \fere,\shiraishi].
This was facilitated  by some new developments in the
the construction of actions for one-dimensional supersymmetric
sigma models [\coles] and the realization that the geometry
of the moduli spaces of black holes will involve a connection
with torsion [\gibbonsp].
The motivation behind this revival is the hope that investigating
these moduli spaces will lead to the understanding of
AdS${}_2$/CFT${}_1$ correspondence and unravel some novel
features in multi-black hole mechanics like bound states, in parallel
with similar developments for BPS monopoles.
The effective theories of the graviphoton  and the
Reissner-Nordstr\"om multi-black holes were constructed in
[\micha,\mss] and the effective theories of all static four- 
and five-dimensional
black holes that preserve four supersymmetry charges were constructed
in [\gutpap, \gutpapb]. For small black hole separations, 
the effective theory
of multi-black holes exhibits  a $D(2,1;0)$ superconformal symmetry 
[\micha,\mss,\gutpapb].
Aspects of the multi-black hole quantum mechanics utilizing
the $D(2,1;0)$ superconformal symmetry have been
investigated in [\strqa, \strqb].

It has been known for sometime that the (4+1)-dimensional black holes 
which preserve four supersymmetries have an $AdS_2\times S^3$ 
near horizon geometry. A
calculation in [\claus, \jade] has revealed that the dynamics of a
probe with fixed angular momentum near a black hole 
with the above near horizon geometry 
is described by de Alfaro, 
Fubini and Furlan conformal (DFF) model [\aff]. The 
coupling of the conformal
model is related to the conserved angular momentum.
Using this, it was later argued in [\gibbonst] 
that the conformal Calogero model
is a candidate for the dual  superconformal theory in the
context of  AdS${}_2$/CFT${}_1$ correspondence.

 One of the results of this
paper is to recover the DFF model from the dynamics
 of multi-black holes
as described by the moduli metric. 
We shall mostly focus on the dynamics of the supersymmetric
black holes of the STU model. For two black holes, we shall show
that, if the center of mass motion decouples
from the relative motion of the system, then the relative 
angular momentum of the pair is conserved.
Further we shall find that the radial relative 
motion of the two black hole
system is described by a sigma model with a scalar potential which
is bell shaped.
In particular, we shall show that for non-vanishing 
relative angular momentum
and depending on the energy and original
separation of the black hole pair,  the two black holes either  become
dynamically well separated or dynamically  approach each other.
In addition, we shall find that for small black hole separations
the effective theory of a black hole pair with non-vanishing relative
angular momentum is given by the DFF model.

We shall also extend the above analysis to systems that involve
more than two black holes. For this, we shall give an explicit
expression for the black hole moduli metric that involves all
the three body interactions. In particular, we shall find that
part of the three body interactions can be computed from the
one-loop three point amplitude of a four-dimensional $\phi^3$-theory.
We shall use several approximations to study the dynamics of three
black holes. In particular, we shall investigate the system
in the limit that two of the black holes are clustered in to a binary
 while the third one is further away. In this approximation, the effective
theory of the system for small black hole separations and
 with non-vanishing
relative angular momentum  is also given by a 
sigma model with
a scalar potential which however is not of Calogero type. 

This paper is organized as follows: In section two, the moduli
metric of a system of  STU black holes is given. In section three,
the dynamics of a system of two STU black holes is investigated.
In section four, the dynamics of a system of more than two black
holes is examined.

\chapter{ Moduli Metric of (4+1)-dimensional Black holes revisited}

It has been found in [\gutpap] that the moduli metric of 
 (4+1)-dimensional
black holes coupled to any number of vectors which
 preserve four supersymmetries
 can be determined from the components of the black hole
metric. Suppose that  the spacetime metric that describes 
$N$ black holes
located at  $\{\fy_A\in \bR^4;A=1, \dots, N\}$ is
$$
ds^2= -A^2(\fx, \fy_A) dt^2+B^2(\fx, \fy_A) ds^2(\bR^4)\ ,
\eqn\aaa
$$
where $t$ is the time coordinate and $\fx\in \bR^4$ are
 the space coordinates. 
Then the metric on the moduli space
is determined by the moduli potential 
$$
\mu(\fy_A)=\int_{\bR^4} d^4x\, A^{-2} B^2 (\fx, \fy_A)
\eqn\modpot
$$
as
$$
ds^2=\big(\partial_{mA}\partial_{nB}+\sum_{r=1}^3 (I_r)^k{}_m 
(I_r)^\ell{}_r
 \partial_{kA}\partial_{\ell B}\big)\mu\  dy^{mA} dy^{nB}\ ,
 \eqn\modmetra
 $$
 where $\{I_r; r=1,2,3\}$ is a constant hypercomplex structure
 on $\bR^4$ associated with a basis of self-dual two-forms
  and $k,\ell,m,n=1,\dots,4$.
 The moduli metric can also be written as
 $$
 ds^2= \big(\delta_{mn} U_{AB}
 + \sum_{s=1}^3\, (V_s)_{AB}\, (J_s)_{mn}\big) dy^{mA} dy^{nB}\ ,
 \eqn\modmetrb
 $$
 where $\{J_s; s=1,2,3\}$ is a constant hypercomplex structure
 on $\bR^4$ associated with a basis of anti-self-dual two-forms and
 $$
 \eqalign{
 U_{AB}&= \delta^{mn}\partial_{mA}\partial_{nB}\mu
 \cr
 \sum_{s=1}^3 V_s (J_s)_{mn}&= \big(\partial_{mA}\partial_{nB}
\mu-(n,m)\big)
 -\epsilon_{mn}{}^{k\ell} \partial_{kA}\partial_{\ell B}\mu \ .}
 \eqn\aab
 $$
 To show this, one uses uses the identity
 $$
 \sum_{r=1}^3 (I_r)^\ell{}_m (I_r)^k{}_n= \delta_{mn} \delta^{\ell k}-
 \delta^k{}_m \delta^\ell{}_n-\epsilon_{mn}{}^{\ell k}
 \eqn\aac
 $$
 and expands the anti-self-dual part of the moduli metric 
in the $J_s$ basis; 
 the spatial
 indices are raised and lowered with respect to the Euclidean metric
 on $\bR^4$. As we shall see the components of the moduli metric
 corresponding to $U_{AB}$ are diagonal in the pair-wise black 
hole separations
 while the terms of the moduli metric proportional to $(V_s)_{AB}$ are
 off-diagonal.

The effective theory
 of black holes associated with moduli metric \modmetra\ has  $N=4B$
 one-dimensional
 supersymmetry [\coles, \gibbonsp]. The 
   black hole moduli space
 is a hyper-K\"ahler manifold with torsion (HKT) [\twist]. 
The  torsion  appears 
 in the couplings of the fermion of the effective theory.

 In what follows, we shall be concerned with the 
dynamics of graviphoton
 and STU model black holes. Since the graviphoton 
black holes are a special
 case of the STU model black holes, we shall
 describe first the moduli
 metric of STU black holes. Then we shall give the limit in which
 the moduli metric of the graviphoton black holes 
arises from the STU ones.
 The moduli potential for the black holes of the  STU model
is 
 $$
 \mu=\int_{\bR^4}\, d^4x \, H_1 H_2 H_3 
 \eqn\modpott
 $$
 where
 $$
 H_i=h_i+\sum_{A=1}^N {\lambda_{iA}\over |\fx-\fy_A|^2}
 \eqn\abc
 $$
 for $i=1,2,3$ which are harmonic functions on $\bR^4$.
 The constants $\{h_i; i=1,2,3\}$ are related to the asymptotic
 values of the two scalars of the theory and the 
constants $\{\lambda_{iA};
 i=1,2,3 ; A=1, \dots, N\}$ are interpreted as the 
charges of the $A$ black-hole
 with respect to the $i$-th Maxwell gauge potential; 
the STU model has three
 Maxwell fields. For the masses of the black holes to be
 positive and for the
 black hole solution not to have naked and other 
singularities, we take
 $h_i,\lambda_{iA}>0$.
 Using the moduli potential \modpott, we find that the 
moduli metric of the
 STU black holes [\gutpap] can be rewritten as 
 $$
 \eqalign{
 ds^2&= V_3 \sum_A[h_2 h_3 \lambda_{1A}+h_1 h_3\lambda_{2A}
 +h_1 h_2 \lambda_{3 A}]
 |d\fy_A|^2
 \cr
+& V_3 \sum_{A\not= B} [h_2 \lambda_{1A} \lambda_{3B}+
h_1 \lambda_{2A} \lambda_{3B}
 +h_3 \lambda_{1A} \lambda_{2B}] { |d\fy_A- d\fy_B|^2\over
 |\fy_A-\fy_B|^2}
 \cr
 +&{V_3\over4} \sum_{ \{ A\not=B \},C} \rho_{ABC}
 |d\fy_A-d\fy_B|^2 
  \big[ {1\over |\fy_A-\fy_C|^2 |\fy_A-\fy_B|^2}
\cr
 & + {1\over |\fy_B-\fy_C|^2 |\fy_A-\fy_B|^2}-{1\over 
|\fy_A-\fy_C|^2 |\fy_B-\fy_C|^2} \big]
  \cr
  -&{2 \over3} \sum_{A\not=B\not=C} \int\, d^4x\, \rho_{ABC} 
   {\big[ (dy_A-dy_C)^{[m} (dy_B-dy_C)^{n]}\big]^-\over 
|\fx-\fy_C|^2} \times
  \cr
 & \qquad \qquad \qquad \qquad \qquad
 \partial_m \big({1\over |\fx-\fy_A|^2}\big) 
 \partial_n \big({1\over |\fx-\fy_B|^2}\big)\ ,}
 \eqn\modstu
 $$
 where $V_3$ is the volume of the unit three
 sphere, $\big[ (dy_A-dy_C)^{[m} (dy_B-dy_C)^{n]}\big]^-$
 denotes the anti-self-dual projection of
 $(dy_A-dy_C)^{[m} (dy_B-dy_C)^{n]}$, and
$$
\rho_{ABC} = \big[ \l_{1A} \l_{2B} \l_{3C} + \l_{1C} \l_{2A} \l_{3B}
+ \l_{1B} \l_{2C} \l_{3A} + (A \leftrightarrow B) \big] \ .
\eqn\rhoeqn
$$
 The moduli metric has a free term for N particles, and 
  two-and three-body velocity dependent interactions.
 Observe that part of the moduli metric that contains 
three body interactions
 is not given explicitly since the last term in \modstu\
 involves an integration over the spatial coordinates
 $\fx$ which has not been carried out.
 The three body interactions in the moduli 
metric that are given explicitly
 are {\it diagonal} in the pair-wise black
 hole separations, $|d\fx_A-d\fy_B|$, 
 while
 the term that involves the integral are {\it off-diagonal}. 
  We remark that if the masses of the black holes 
 $m_A=V_3(h_2 h_3 \lambda_{1A}+h_1 h_3\lambda_{2A} 
+h_1 h_2 \lambda_{3 A})$
 are not equal, then the centre of mass motion does not decouple
 from the dynamics of the system.

 The graviphoton black holes arise in the case where $H_1=H_2=H_3=H$.
 The presence of one independent harmonic function
 implies that the two scalars of the STU black hole solutions 
  are constant. Moreover the black holes are charged with respect to
 a single Maxwell gauge potential which is a linear 
combination of the three 
 gauge potentials of the STU model. This single gauge potential
 is identified with the graviphoton of the simple (4+1)-dimensional
 supergravity. The moduli metric of the graviphoton 
black holes [\micha]
 is given from that of the STU model black  holes \modstu\ by setting
 $h_1=h_2=h_3=h$ and $\lambda_1=\lambda_2=\lambda_3=\lambda$.

 At small black hole separations, the dynamics of 
the system is dominated
 by the three body interactions. The moduli metric
 in this limit is given as in \modstu\ but with $h_i=0$ for $i=1,2,3$.
 The effective theory of supersymmetric black holes apart from the
 kinetic term involving the moduli metric, it also contains various
 fermionic terms. In what follows, we shall focus on 
the classical dynamics
 governed by the moduli metric only.

\chapter{Two Black Hole System}

\section{Moduli Metric}
The moduli metric for two black holes can be written as
$$
\eqalign{
ds^2&= m_1 |d\fy_1|^2+ m_2 |d\fy_2|^2+ g_{(2)}
{|d\fy_1-d\fy_2|^2\over |\fy_1-\fy_2|^2}
\cr &+g_{(3)}
{|d\fy_1-d\fy_2|^2\over |\fy_1-\fy_2|^4}\ ,}
\eqn\bba
$$
where
$$
\eqalign{
m_1&=V_3[h_2 h_3 \lambda_{11}+h_1h_3 \lambda_{21} +
h_1 h_2 \lambda_{31}]
\cr
m_2&=V_3[h_2 h_3 \lambda_{12}+h_1h_3 \lambda_{22} +
h_1 h_2 \lambda_{32}]
\cr
g_{(2)}&= V_3[ h_2\lambda_{11} \lambda_{32}+
h_1 \lambda_{21}\lambda_{32}+
h_3 \lambda_{11}\lambda_{22}+ h_2\lambda_{12} \lambda_{31}
+h_1 \lambda_{22}\lambda_{31}+
h_3 \lambda_{12}\lambda_{21}]
\cr
g_{(3)}&= {V_3 \over 2} (\rho_{122}+ \rho_{211}) \ .}
\eqn\bbc
$$
Using the conditions that we have put on the
 parameters of the classical
solution, we find that $m_1, m_2, g_{(2)}, g_{(3)}>0$.
In particular for the graviphoton black holes, we have
$$
\eqalign{
m&=3~V_3~h^2~\lambda
\cr
g_{(2)}&=6~ V_3~ h~ \lambda^2
\cr
g_{(3)}&= 6~V_3~\lambda^3\ .}
\eqn\gravcoupl
$$

The term that contains the integral in \modstu\   does 
not contribute in this case.
In the limit of small separations, the moduli
metric becomes
$$
ds^2=g_{(3)}
{|d\fy_1-d\fy_2|^2\over |\fy_1-\fy_2|^4}\ .
\eqn\bcb
$$

A direct observation reveals that if the masses $m_1, m_2$ of the two
black holes are different, then the centre of mass motion
of the two black holes does {\it not} decouple from the relative motion
of the system. For the graviphoton black holes
the centre of mass motion decouples as well as for those STU black holes
for which  $\lambda_{i 1}
=\lambda_{i 2}$ for $i=1,2,3$. However for {\it generic} STU black holes,
the centre of mass motion does {\it not} decouple. 

Suppose that $m=m_1=m_2$ and the center of mass motion decouples
from the relative one. In this case we set $\fr={\fy_1-\fy_2\over2}$
and $\fu={\fy_1+\fy_2\over2}$  where $\fu$ is the position of the
center of mass and $\fr$ is the relative position of the two black holes. 
The moduli metric for such a two black
hole system can be rewritten as
$$
ds^2= 2m |d\fu|^2+2m |d\fr|^2+ g_{(2)} {|d\fr|^2\over |\fr|^2}+4 g_{(3)} 
 {|d\fr|^2\over |\fr|^4}\ .
 \eqn\bbd
$$

\section{Dynamics at Fixed Angular Momentum}

The moduli metric of the two black hole system is invariant under
all $SO(4)$ rotations acting on the positions of the black holes
with infinitesimal transformations
$$
\delta \fy_A= \omega \fy_A\ ,
\eqn\caa
$$
where $\omega$ is a $4\times 4$ skew-symmetric matrix.
For those black hole systems for which the relative motion
decouples for the centre of mass motion, the part of the 
moduli metric which describes the relative motion is also
invariant under the $SO(4)$ rotational symmetry
$\delta \fr= \omega \fr$.
This is the case on which we shall focus our attention.

Changing coordinates from Euclidean to angular, 
we write the relative
moduli metric as
$$
ds^2= \phi^2(r) (ds^2+ r^2 ds^2(S^3))
\eqn\ccc
$$
where $ds^2(S^3)$ can be parameterized with respect to the right-
invariant one forms $\sigma^i$,
 $ds^2(S^3)=\sum_{i=1}^3 (\sigma^i)^2$,
where $d\sigma^i=-\epsilon^i{}_{jk} \sigma^j\wedge \sigma^k$
and
$$
\phi^2(r)= 2m+{g_{(2)}\over r^2}+{4g_{(3)}\over r^4}\ .
\eqn\ccd
$$
Observe that since $m_1, m_2, g_{(2)}, g_{(3)}>0$, then $\phi^2>0$.

To be explicit we write the metric on the relative
 displacement 3-sphere as
$$
ds^2(S^3) = d \t^2 +\sin^2 \theta \big( d \b^2 +
 \sin^2 \b d \a^2 \big)
\eqn\cce
$$
where the relative displacement co-ordinates are
$$
{\bf r} = \pmatrix{ r \sin \t \sin \b \sin \a \cr r \sin \t 
\sin \b \cos \a \cr r \sin \t \cos \b \cr r \cos \t}
\eqn\ccf
$$
and the right invariant 1-forms are
$$
\eqalign{
\s^1 = & - \sin \b \cos \a d \t + \sin \t \sin \b (\sin \a \cos \t 
+ \sin \t \cos \a \cos \b) d \a 
\cr 
&- \sin \t (-\sin \a \sin \t + \cos \b \cos \a \cos \t) d \b
\cr
\s^2 = &- \sin \b \sin \a d \t - \sin \t \sin \b 
(-\sin \t \cos \b \sin \a 
+ \cos \a \cos \t) d \a
\cr  
&+ \sin \t (-\sin \t \cos \a - \cos \t \cos \b \sin \a) d \b
\cr
\s^3 = &- \cos \b d \t + \sin \b \sin \t (\cos \t d \b - 
\sin \t \sin \b d \a)\ .}
\eqn\cct
$$
The conserved angular momentum associated with the $SO(3)$ 
subgroup of $SO(4)$
that leaves $\sigma^i$ invariant is
$$
J^i=\phi^2  r^2 \sdt^i\ ,
\eqn\ccs
$$
where now $\sdt^i dt$ is the pull-back of $\sigma^i$ on
 the worldline and
$J^i$ is conserved, ${d\over dt} J^i=0$. 
The equation  for the radial motion is
$$
-{d\over dt} [\phi^2  {d\over dt} r]+{J^2\over r^3 \phi^2}+ 
\partial_r\phi^2 \big(({d\over dt} r)^2+ {J^2\over r^2\phi^4}\big)=0\ .
\eqn\eqmot
$$
The equations of motion associated with the angular coordinates
simply imply the conservation of angular momentum.
The equation of motion \eqmot\ is that of a non-relativistic 
particle with Lagrangian
$$
L={1\over2} \phi^2 ({d\over dt} r)^2- V(r)\ ,
\eqn\lag
$$
where
$$
V(r)={J^2\over 2} {1\over r^2 \phi^2(r)}\ .
\eqn\dft
$$
In fact, this potential is associated with the
 superpotential $W=\log r$ as
$$
V(r)={J^2\over 2} \phi^{-2} (\partial_rW)^2\ ,
\eqn\potpot
$$
as it may have been expected because the original theory
 is supersymmetric.

The energy of the system associated with the Lagrangian \lag\ is
$$
E={1\over2} \phi^2 ({d\over dt} r)^2+{J^2\over 2} {1\over r^2 \phi^2(r)}
\eqn\haa
$$
and it is conserved.

There are two cases to consider regarding the dynamics of the system
with potential \potpot. First if $J^2=0$, then  $V$ vanishes.
The two black holes can be located at any point in $\bR^4$
 provided that $E=0$.
The relative dynamics of the system in this case is determined
by solving the equation
$$
{d\over dt} r=\pm {\sqrt{2 E}\over \phi}\ .
\eqn\hba
$$

Next suppose that $J^2\not=0$. The potential $V(r)>0$
 and $V(r)\rightarrow 0$ as $r\rightarrow 0$ and 
 $r\rightarrow +\infty$. The potential $V(r)$ has  
two  critical points
 on the positive real line at $r=0$ and 
$r^2_*= (2g_{(3)}/m)^{1\over2}$, as  
 can be easily seen by computing $\partial_r V=0$, ie
 $V(r)$ has a bell shape. The critical
 point at $r=r_*>0$ is a global maximum. 
 The value of the potential at the maximum  is
 $V_*=V(r_*)={J^2\over\ 8 \sqrt{2 m g_{(3)}+ 2g_{(2)}}}$.
 If the  black holes with relative angular momentum  $J^2\not=0$
  are separated by less than the critical distance
 $r<r_*$ and have energy $E<V_*$, then they roll down the 
 potential towards $r\rightarrow 0$
 and so their separation is  dynamically reduced
 \foot{We remark that if the initial separation is close to the
 critical distance then by the time that 
the black hole have approached
 $r\rightarrow 0$, the non-relativistic approximation may not be
 valid.}.
 On the other hand if the black holes are separated by $r>r_*$ and 
 have energy $E<V_*$, then
 they will again roll down the other side of the potential towards
 $r\rightarrow +\infty$. In this case the two black
 holes become dynamically
 well separated. 
 
 The equation of motion for black holes with $J^2\not=0$ is
 $$
 {d\over dt} r=\pm {1\over \phi(r)} {\sqrt{2E-
 {J^2\over r^2 \phi^2}}}\ .
 \eqn\hhcd
 $$ 
 In the limit of small black hole separations, the dynamics along
the radial direction simplifies considerably. In this case
$\phi^2=4 g_{(3)}/r^4$. Before one proceeds with the calculation above,
it is best to change the radial coordinate 
as $q=2 \sqrt{ g_{(3)}}/ r$.
 In this case,  the moduli metric becomes
 $ds^2=dq^2+q^2 ds^2(S^3)$, ie the standard Euclidean metric
 on $\bR^4$ but in angular coordinates.

Focusing on the dynamics of the two black hole system
with non-vanishing relative angular momentum, we find that the
associated Lagrangian which describes the radial motion is     
$$
L={1\over2} [({d\over dt} q)^2 -{J^2\over q^2}]\ .
\eqn\hggy
$$
This is precisely the  Lagrangian of the DFF model. Therefore
we have shown that the  probe computation of [\claus, \jade] coincides
with the result obtained from the moduli approach.

\chapter{Black Hole Three Body Interactions }

\section{Three body interactions and $\phi^3$ theory}
 
To make progress towards an explicit expression for the
moduli metric of $N>2$ (4+1)-dimensional black holes, 
we have to evaluate
the integral in \modstu\ which involves the off-diagonal 
 three body interactions.
This contribution to the moduli metric can be rewritten as
$$
\eqalign{
ds^2_{OD}&=-{2\over3}
\sum_{A\not=B\not=C} \rho_{ABC}
 \big[ (dy_A-dy_C)^{[m} (dy_B-dy_C)^{n]}\big]^- 
\partial_{mA}\partial_{nB} 
  {\cal A}(\fy_A, \fy_B, \fy_C)\ ,}
  \eqn\jjuu
$$
where
$$
{\cal A}(\fy_A, \fy_B, \fy_C)=\int\, d^4x\, 
  {1\over |\fx-\fy_C|^2} 
  {1\over |\fx-\fy_A|^2} 
  {1\over |\fx-\fy_B|^2}\ .
  \eqn\uuy
$$
Observe that since the positions  $\fy_A$, $\fy_B$, 
$\fy_C$ are all distinct,
${\cal A}$ is finite.
It can be immediately recognized that ${\cal A}$ is the one-loop
three point green function of a massless $\phi^3$ theory. For
this, $\fx$ is identified as the loop momentum $\fk$ and 
$\fp_{AB}= \fy_A-\fy_B$
are identified
as the three incoming momenta. To compute this integral, we shall use
field theory techniques and terminology. After changing variables, 
${\cal A}$ can be written as
$$
{\cal A}(\fp_{AC}, \fp_{BC})=\int\, d^4k\, 
  {1\over |\fk|^2} 
  {1\over |\fk-\fp_{AC}|^2} 
  {1\over |\fk-\fp_{BC}|^2}\ .
  \eqn\vvt
$$
Applying standard field theory methods,  ${\cal A}$ can be expressed as
$$
{\cal A}=2\int\, d^4k\ \int_0^1\, d\alpha\,\int_0^\alpha\, d\beta
[|\fk-\fp_{BC}|^2 \beta+(\alpha-\beta) |\fk-\fp_{AC}|^2
+|\fk|^2(1-\alpha)]^{-3}\ .
\eqn\rrt
$$
After again changing variables 
$\fk\rightarrow \fk-\fp_{BC} \beta-\fp_{AC} (\alpha-\beta)$ 
and performing
the integration over $\fk$, we find that 
 $$
{\cal A}= {V_3\over2}
\int^1_0 d\alpha \int^\alpha_0 d\beta [ |\fp_{AB}|^2 (\alpha-\beta) \beta
+(1-\alpha)
 (\beta |\fp_{BC}|^2+(\alpha-\beta) |\fp_{AC}|^2)]^{-1}\ . 
 \eqn\sstr
$$
It is next convenient to change co-ordinates by
 defining $\b = \a w$, and to set $m=0$. Then the 
integral may be rewritten as
$$
{\cal A}= {V_3\over2}
\int^1_0 \int^1_0 d \a dw {1 \over |\fp_{AB}|^2 \a w(1-w) 
+(1-\a)(|\fp_{BC}|^2 w +(1-w) |\fp_{AC}|^2)}\ .
\eqn\yytru
$$
Carrying out the $\a$-integral, we obtain
$$
{\cal A}=- {V_3\over2}
\int^1_0 dw {\log \big({ |\fp_{AB}|^2 w(1-w) \over |\fp_{BC}|^2 w 
+ |\fp_{AC}|^2(1-w)} \big)
\over \big[ |\fp_{AB}|^2 w^2 + (|\fp_{BC}|^2 
- |\fp_{AC}|^2 -|\fp_{AB}|^2)w + |\fp_{AC}|^2 \big] }\ .
\eqn\nthh
$$
The full expression for this 
integrand is given in the appendix. 
As it is difficult to obtain an insight of the black hole
dynamics from the resulting rather
 complicated expression, we shall use 
an approximation to investigate some of its properties.

\section{ Black Hole Binaries}

In the black hole binary approximation, black holes 
$A$ and $B$ are 
close together while the  black hole $C$ is further way.
 In such an approximation,
$\fab^2 \ll \fac^2$ and $\fab^2 \ll \fbc^2$. This implies that
$ | \fbc^2 - \fac^2| \ll \fac^2$ and also $ |\fbc^2 - \fac^2 | 
\ll \fbc^2$. It is also
useful to change co-ordinates again by setting $w={1 \over 2}+v$. Then
the symmetry of ${\cal{A}}$ under the interchange
 $A \leftrightarrow B$  is made manifest as
$$
{\cal A} = -{V_3 \over 2} \int_{-{1 \over 2}}^{1 \over 2} dv 
{\log \big( { \fab^2({1 \over 4}-v^2) \over v (\fbc^2- \fac^2)
+{1 \over 2}(\fbc^2 + \fac^2)} \big) \over (v^2-{1 \over 4}) \fab^2 
+ v (\fbc^2-\fac^2) +{1 \over 2} (\fac^2+\fbc^2)}\ .
\eqn\dftrc
$$
It is clear that under the above assumptions
 ${\fab^2 \over \fac^2+ \fbc^2}$ is a small dimensionless parameter
 which we can use to expand the integral out. In particular we find
 that
$$
\eqalign{
{\cal A} &= -{V_3 \over 2}  \int_{-{1 \over 2}}^{1 \over 2} 
dv {2 \over \fac^2 + \fbc^2}
\times 
\cr
&\big[ \big[1 - {2 \over \fac^2 + \fbc^2} \big((v^2-{1 \over 4}) 
\fab^2 
+ v (\fbc^2 - \fac^2) \big)
\big]
\times
\cr
&\big[ \log \big( {2 \fab^2 ({1 \over 4}-v^2) \over \fbc^2 
+ \fac^2} \big) - {2 v (\fbc^2 - \fac^2) \over \fac^2 
+ \fbc^2} \big] 
+ O \big( \big({\fab^2 \over \fac^2+ \fbc^2}\big)^2 \big)
 \big]\ . }
\eqn\potry
$$
The leading order term in this expression is given by
$$
{\cal A} = - V_3 {\log \big( {\fab^2 \over \fbc^2+ \fac^2} \big)
 \over \fbc^2
+ \fac^2}\ .
\eqn\qrtew
$$
Acting on this term with the differential operator 
$\pd{mA} \pd{nB} - \pd{nA} \pd{mB}$,
we obtain from this the  leading order off-diagonal 
contribution to the relevant component of the metric which is
given by
$$
\eqalign{
ds^2_{OD}&\sim-{8 V_3\over3} \rho_{ABC} 
 {(y_A-y_B)_m (y_A+y_B-2y_C)_n 
  \over \fab^2 
(\fac^2 + \fbc^2)^2} \big[ (dy_A-dy_C)^{[m} (dy_B-dy_C)^{n]}\big]^-
  \ ,}
  \eqn\sxtr
$$
where $\sim$ indicates that the moduli metric is corrected by higher
order terms.
We observe that this expression is of order ${1 \over \fab \fac^3}$. 
The remaining  3-body interactions
$$
\eqalign{
ds^2_{D}=& {V_3\over4} \sum_{ \{ D\not=E \},F} \rho_{DEF}
 |d\fy_D-d\fy_E|^2 
  \big[ {1\over |\fy_D-\fy_F|^2 |\fy_D-\fy_E|^2}
\cr
 & + {1\over |\fy_E-\fy_F|^2 |\fy_D-\fy_E|^2}-{1\over 
|\fy_D-\fy_F|^2 |\fy_E-\fy_F|^2} \big]\ ,}
  \eqn\xxtr
$$
which are diagonal in the separations,
    contain  terms 
  of order ${1 \over \fab^4 }$,
 ${1 \over \fab^2 \fac^2}$ and ${1 \over  \fac^4}$. 
 So, comparing the components of $ds^2_{OD}$ and $ds^2_{D}$,
   it is consistent to consider two different 
approximations for the moduli metric. 
First, we can keep the components of $ds^2_{D}$ which are of
 order ${1 \over \fab^4 }$ and
 ${1 \over \fab^2 \fac^2}$, and the leading order term in $ds^2_{OD}$.
 Alternatively, we may retain only the components of $ds^2_{D}$
 which are of order ${1 \over \fab^4 }$ and
 ${1 \over \fab^2 \fac^2}$.

Clearly the approximation that we are considering applies for a system
of  three-black
holes for which two of them are clustered together 
to form a binary system while
the third is further way. For more than three black holes, the system is
more complicated, as three or more black holes can cluster together
and so the associated full three-body interaction becomes relevant.
However one can envisage  the possibility where the black holes cluster
in binaries which have separations larger than those of the 
black holes within the binaries. In such a case the 
above approximation applies
for a system of more than three black holes.

\section{Angular Momentum and Black Hole Binaries}

We can now investigate whether in the black hole binary approximation
it is possible to carry out an analysis similar to that 
which we have performed
in section three for the two black hole system.
One of the difficulties which emerges is that, although the angular
momentum of the whole system is conserved, the angular
momentum of each black hole or black hole pair  is not.
To simplify the analysis, we shall consider 
the case of three black holes.
First, we shall  use
the approximation of the previous section in which
 only the components in the
moduli metric diagonal in the black hole separations contribute.

Now suppose that the black holes $A=1$ and $B=2$ are close together
while the black hole $C=3$ is further away. So if we set 
$\fv=\fy_1-\fy_2$
and $\fw=\fy_1-\fy_3$, we have that $|\fv|<<|\fw|$ and
 $\fw\sim \fy_2-\fy_3$.
Working in the limit in which  the three-body interactions dominate
and  off-diagonal components
in the separations are neglected, we find that the moduli metric is
$$
ds^2\sim f^2(|\fv|, |\fw|) | d\fv|^2
\eqn\eeetr
$$
where
$$
f^2(|\fv|, |\fw|)={V_3\over2} 
\big({\rho_{122}+\rho_{211}\over |\fv|^4}
+{2 \rho_{123}\over |\fv|^2 |\fw|^2}\big)\ .
\eqn\qqqtr
$$
In this approximation the angular momentum of the pair of 
the $A=1$ and $B=2$
black holes is conserved. So we can now use the analysis
 we have done in section
three to investigate the effective potential associated 
with the system
for fixed angular momentum.  We find that the 
associated potential is
$$
V(q_1, q_2)={J^2\over V_3 \big[ (\rho_{122}+\rho_{211})
 |q_1|^2+ 2\rho_{123}|q_2|^2 \big]}
\eqn\mmny
$$
where $q_1=|\fv|^{-1}$ and $q_2=|\fw|^{-1}$. Observe 
that this potential
is not of the conformal Calogero type.

Next consider the approximation where the leading  
term in $ds^2_{OD}$ contributes as well. In such a case 
the moduli metric
for the three black hole system is
$$
ds^2\sim f^2(|\fv|, |\fw|) | d\fv|^2- {4 \rho_{123} \over
 3 |\fv|^2 |\fw|^4} (\fv\wedge \fw)\cdot
(d\fv\wedge d\fw)^-\ ,
\eqn\nnbtxt
$$
using a self-explanatory notation. It is clear from this that the
addition of the leading order term in $ds^2_{OD}$ violates the
conservation of the angular momentum of each pair of 
black holes in the system.
To conclude, we remark that it is curious that the four-dimensional
theory with cubic interactions enters in the 
investigation of the moduli metric
for (4+1)-dimensional black holes. It is not clear whether this
is simply a technical coincidence or if  there is a more fundamental
reason for it.

\vskip 1.0cm
\noindent{\bf Acknowledgments:} We thank Alexis Polychronakos
for helpful discussions. 
G.P. is supported by a University Research
Fellowship from the Royal Society. J.G. is supported 
by a EPSRC postdoctoral grant.
This work is partially supported by SPG grant PPA/G/S/1998/00613.

\Appendix{:\quad The three-body interaction moduli potential}

Here we shall give the expression for the amplitude
${\cal A}$. For this we take \dftrc\ and make the change
of variables $v={z \over 2}$. It is also convenient to
define $\g ={1 \over 2} \fab^2$, $\tau = \fbc^2 - \fac^2$, 
$\s = \fbc^2 + \fac^2$. Then  
$$
{\cal{A}} = -{V_3 \over 2} \int_{-1}^1 dz {1 \over \g z^2 + 
\tau z + \s - \g} \log
 \big({\g (1-z^2) \over \tau z + \s} \big)
$$
and from the definitions of $\g$, $\tau$ and $\s$ it follows that
$|\tau|<2 \g$, $2 \g < \s$ and $\s > |\tau|$.
From these inequalities we observe that $ \g z^2 + 
\tau z + \s - \g>0$
for $z \in \bR$.
Defining $a^\pm = -{\tau \over 2 \g} \pm i \sqrt{{\s \over \g}
 - {\tau^2 \over 4 \g^2} -1}$, we have $ \g z^2 + \tau z + \s 
- \g = \g(z-a^+)(z-a^-)$.
Hence
$$
{\cal{A}} = -{V_3 \over 2 \g (a^+ - a^-)} \int_{-1}^1 dz
\big({1 \over z-a^+} - {1 \over z-a^-} \big) \log
 \big({\g (1-z^2) \over \tau z + \s} \big) \ .
$$
Evaluating this expression we obtain
$$
\eqalign{
{\cal{A}}   = &  -  {V_3 \over 2 \g (a^+ - a^-)}  \times 
\cr
& \big[ 
-\dilog ({2 \over 1-a^+}) -  \dilog ({2 \over 1+a^+})
- \dilog \big( { {\s \over |\tau|} +1 \over a^+ +  {\s \over
|\tau|}} \big) +  \dilog \big( { {\s \over |\tau|} -1 \over a^+ 
+  {\s \over
|\tau|}} \big)
\cr
&+\log {\g \over |\tau|} \big( \log(a^+ +1) -\log(a^+-1) \big)
 + \log \big( {a^+ +1 \over a^+ -1} \big) \big( \log(a^+ +1)
 -\log(1-a^+) \big)
\cr
&+ \log (a^+ + {\s \over |\tau|}) \big( \log \big( {a^+ +1 \over a^+ 
+ {\s \over
|\tau|}} \big) -  \log \big( {a^+ -1 \over a^+ + {\s \over
|\tau|}} \big) \big) - (a^+ \leftrightarrow a^-)  \big] \ ,}
$$
where $\dilog$ denotes the principal branch of the dilogarithm function
defined on $\bC$ cut along $\Re \ (x) = (-\infty , 0)$, and
$$
\dilog  \ x  = \int_1^x dt  \ {\log t \over 1-t} \ .
$$
\refout

\end